\documentclass[preprint,superscriptaddress]{revtex4-1}

\usepackage{soul}
\usepackage{ulem}

\usepackage{graphicx}
\usepackage{epstopdf}
\usepackage[ansinew]{inputenc}
\usepackage{array}
\usepackage{color}
\usepackage{amsmath}
\usepackage{amsxtra}
\usepackage{amstext}
\usepackage{amssymb}
\usepackage{latexsym}
\usepackage{dsfont}

\begin{document}
\renewcommand{\figurename}{Figure}

\title{Electric Field Control of Soliton Motion and Stacking in Trilayer Graphene}

\author{Matthew Yankowitz}
\affiliation{Physics Department, University of Arizona, Tucson, AZ 85721, USA}
\author{Joel I-Jan Wang}
\affiliation{Department of Physics, Massachusetts Institute of Technology, Cambridge, MA 02138, USA}
\affiliation{School of Engineering and Applied Sciences, Harvard University, Cambridge, MA 02138, USA}
\author{A. Glen Birdwell}
\affiliation{Sensors and Electron Devices Directorate, US Army Research Laboratory, Adelphi, MD 20783, USA}
\author{Yu-An Chen}
\affiliation{Department of Physics, Massachusetts Institute of Technology, Cambridge, MA 02138, USA}
\author{K. Watanabe}
\author{T. Taniguchi}
\affiliation{National Institute for Materials Science, 1-1 Namiki, Tsukuba 305-0044, Japan}
\author{Philippe Jacquod}
\affiliation{Physics Department, University of Arizona, Tucson, AZ 85721, USA}
\author{Pablo San-Jose}
\affiliation{Instituto de Ciencia de Materiales de Madrid (ICMM-CSIC), Cantoblanco, 28049 Madrid, Spain}
\author{Pablo Jarillo-Herrero}
\affiliation{Department of Physics, Massachusetts Institute of Technology, Cambridge, MA 02138, USA}
\author{Brian J. LeRoy}
\email{leroy@physics.arizona.edu}
\affiliation{Physics Department, University of Arizona, Tucson, AZ 85721, USA}
\date{\today}

\begin{abstract}

The crystal structure of a material plays an important role in determining its electronic properties.  Changing from one crystal structure to another involves a phase transition which is usually controlled by a state variable such as temperature or pressure.  In the case of trilayer graphene, there are two common stacking configurations (Bernal and rhombohedral) which exhibit very different electronic properties~\cite{Aoki2007,Guinea2006,Peeters2009a,Koshino2009,Peeters2009b,Peeters2010,Koshino2010,MacDonald2010,Kumar2011,Wu2011,Tang2011}.  In graphene flakes with both stacking configurations, the region between them consists of a localized strain soliton where the carbon atoms of one graphene layer shift by the carbon-carbon bond distance ~\cite{Zhang2013,Vaezi2013,SanJose2013,Xu2012,Warner2012,Hattendorf2013,Alden2013}.  Here we show the ability to move this strain soliton with a perpendicular electric field and hence control the stacking configuration of trilayer graphene with only an external voltage.  Moreover, we find that the free energy difference between the two stacking configurations scales quadratically with electric field, and thus rhombohedral stacking is favored as the electric field increases.  This ability to control the stacking order in graphene opens the way to novel devices which combine structural and electrical properties.
\end{abstract}

\maketitle

Multilayer graphene has attracted interest in large part due to the ability to induce a sizable band gap with the application of an electric field.  The exact nature of the electronic properties of multilayer graphene is controlled both by the number of layers as well as their stacking configuration.  The equilibrium in-plane crystal structure of graphene is hexagonal ~\cite{CastroNeto2009}, and deviations from this equilibrium require a large amount of energy.  Upon stacking multiple graphene sheets, Bernal-stacking -- where the A-sublattice of one layer resides above the B-sublattice of the other layer -- represents the lowest energy stacking configuration.  Thus under normal circumstances, any two graphene layers in a graphite stack will be Bernal-stacked with respect to one another.  However, when examining layers more than one apart, there can be multiple nearly-degenerate stacking configurations ($2^{(n-2)}$ such configurations for {\it n} layers) ~\cite{Aoki2007}.  For example, in the simplest case of trilayer graphene, the top layer may lie directly above the bottom layer (denoted Bernal- or ABA-stacked), or may instead be configured such that one sublattice of the top layer lies above the center of the hexagon of the bottom layer (denoted rhombohedrally- or ABC-stacked).  Applying a perpendicular electric field breaks the sublattice symmetry differently depending on the stacking configuration, and thus is capable of re-ordering the energy hierarchy of the stacking configurations ~\cite{Aoki2007,Guinea2006,Peeters2009a,Koshino2009,Peeters2009b,Peeters2010,Koshino2010,MacDonald2010,Kumar2011,Wu2011,Tang2011}.  As a consequence, multilayer graphene exhibits the rare behavior of crystal structure modification, and hence modification of electronic properties, via the application of an external electric field.

\begin{figure}[t]
\includegraphics[width=8.5cm]{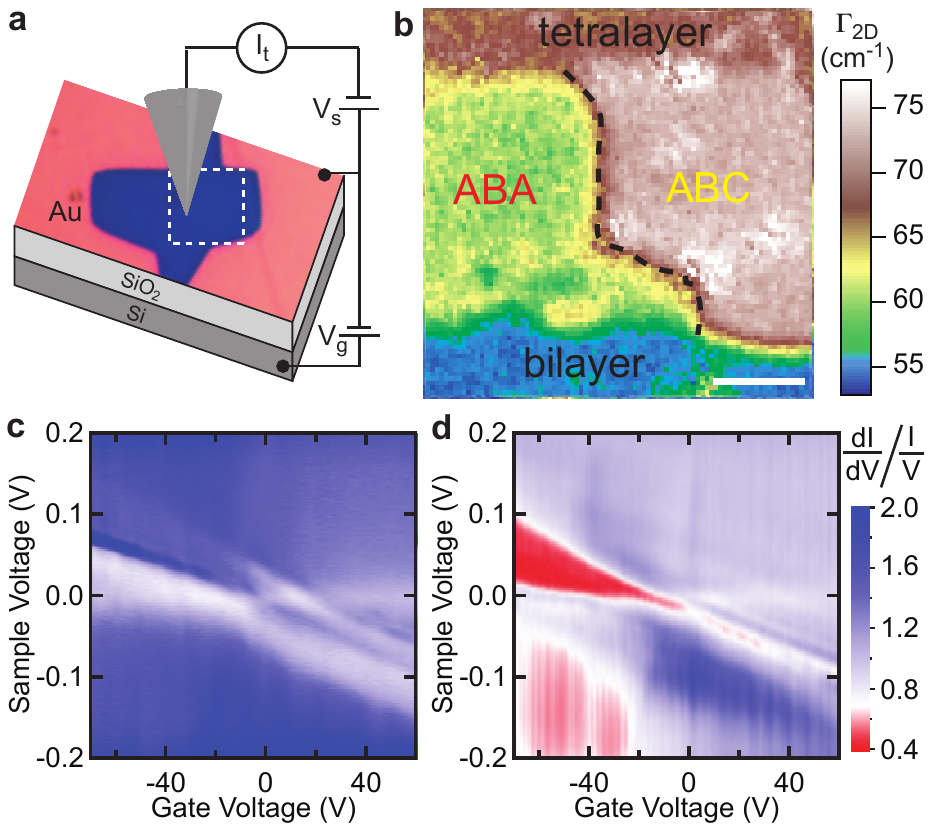} 
\caption{Experimental setup and trilayer graphene spectroscopy. (a) Schematic of the measurement setup showing the STM tip and an optical microscope image of the measured sample.  The dotted white box denotes the region shown in (b).  (b) Raman mapping of the graphene flake on hBN showing areas of bilayer, trilayer and tetralayer graphene.  In the trilayer region, the regions of different stacking order are marked.  The measured soliton is denoted by the dashed black line.  The scale bar is 1 micron.  (c) and (d) Normalized differential conductance (dI/dV)/(I/V) as a function of sample voltage and gate voltage in the ABA and ABC regions, respectively.  A band gap opens at large gate voltages in the ABC region.  For both measurements the current was stabilized at 100 pA at 0.2 V.}
\newpage
\label{fig:schematic}
\end{figure}

To examine this effect, we perform scanning tunneling topography (STM) and scanning tunneling spectroscopy (STS) measurements of trilayer graphene on hexagonal boron nitride (hBN).  Fig. ~\ref{fig:schematic}(a) shows a schematic of our experimental setup with the STM tip and the electrical connections indicated.  Fig. ~\ref{fig:schematic}(b) shows Raman spectroscopy mapping of the graphene on hBN flake measured in this study.  A central region of trilayer graphene is surrounded by a bilayer region below and a tetralayer region above.  The left side of the trilayer region is ABA-stacked and the right side is ABC-stacked.  These regions are identified by a change in the width of the Raman 2D peak ~\cite{Lui2011, Cong2011}.  A smooth transition of the stacking order can be achieved via a domain wall with a localized region of strain (a strain soliton), where one layer shifts by the carbon-carbon spacing, $a_0 = 1.42$ \AA ~\cite{Zhang2013,Vaezi2013,SanJose2013,Xu2012,Warner2012,Hattendorf2013,Alden2013}.  The interface lies above a flat region of hBN, is atomically smooth in STM topography measurements, and does not display a sizable moir\'e pattern \cite{Yankowitz2012}; therefore it is a good candidate for the study of the intrinsic physics of the domain wall.  The ends of the domain wall are bounded by the bilayer and tetralayer regions.

%%%%%%%%%%%%%%%
%%%%%%%%%%%%%%%%%%

An STM tip is used to scan across the domain wall separating the ABA- and ABC-stacked trilayer graphene regions.  There is a net electric field in the region underneath the tip created by voltage differences between the STM tip, silicon back gate and graphene.  Figs. ~\ref{fig:schematic}(c) and (d) show normalized (dI/dV)/(I/V) spectroscopy as a function of gate voltage for the ABA and ABC stacking orders respectively, taken far from the domain wall.  The results are similar to those seen in trilayer graphene on SiO$_2$ \cite{Yankowitz2013}.  Most importantly, the ABA region remains metallic for all gate voltages probed.  In contrast, a sizable band gap can be opened in the ABC region with the application of large gate voltages.  The spectroscopy for the two stacking orders is easily distinguishable for all gate voltages, even within a few nanometers of the domain wall separating the two stacking orders.  This permits very accurate determination of the domain wall location using spectroscopy measurements.

\begin{figure}[t]
\includegraphics[width=8.5cm]{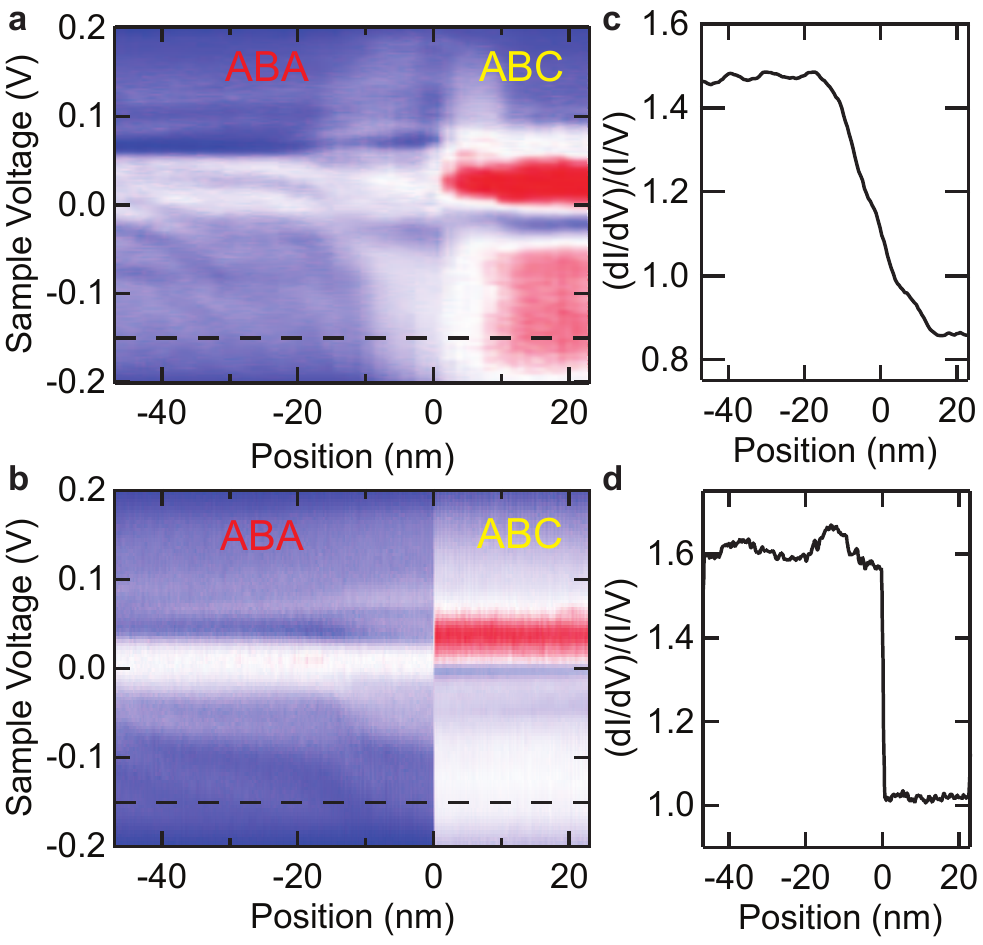} 
\caption{Spatially resolved spectroscopy across a domain wall separating ABA and ABC stacking. (a) Normalized (dI/dV)/(I/V) spectroscopy as a function of tip position and sample voltage for a pinned domain wall.  The tip is moving from left to right and the domain wall appears with a width of $\sim$20 nm.  (b) Normalized (dI/dV)/(I/V) spectroscopy as a function of tip position and sample voltage for a free domain wall.  The tip is moving from left to right and the spectroscopy abruptly changes from ABA to ABC.  For both measurements the current was stabilized at 100 pA at 0.2 V.  The data was acquired with a large negative voltage on the back gate. (c) and (d) Line cuts of (a) and (b), respectively, at a fixed sample voltage of -150 mV (indicated by black dotted lines in (a) and (b)).}
\label{fig:linespectroscopy}
\newpage
\end{figure}

To investigate the connection between the position of the domain wall and the electronic properties of trilayer graphene, we perform dI/dV spectroscopy as a function of tip position scanning from the ABA to ABC region.  Figs. ~\ref{fig:linespectroscopy}(a) and (b) show two examples of these measurements (normalized by I/V), taken at different locations on the soliton and at large negative gate voltages (where there is a large gap in the ABC trilayer region).  Fig. ~\ref{fig:linespectroscopy}(a)  is taken within a few hundred nanometers of the bilayer edge.  In this case, the spectroscopy smoothly evolves from ABA to ABC over a spatial extent of about $\sim$20 nm.  Fig. ~\ref{fig:linespectroscopy}(b) is taken closer to the center of the trilayer region.  In this case, the spectroscopy abruptly changes from ABA to ABC.  In this region of the sample, even maps with atomic resolution show an abrupt transition from ABA to ABC.  As we argue below, this peculiar behavior is due to the STM tip dragging the domain wall for a finite distance along the sample before it snaps back to its equilibrium position.  For the case of Fig. ~\ref{fig:linespectroscopy}(a), the STM tip is very close to the pinned boundary (the bilayer edge) and therefore the energy cost of moving the domain wall is too large to overcome.

%%%%%%%%%%%%%%%
%%%%%%%%%%%%%%%%%%

\begin{figure}[h]
\includegraphics[width=8.5cm]{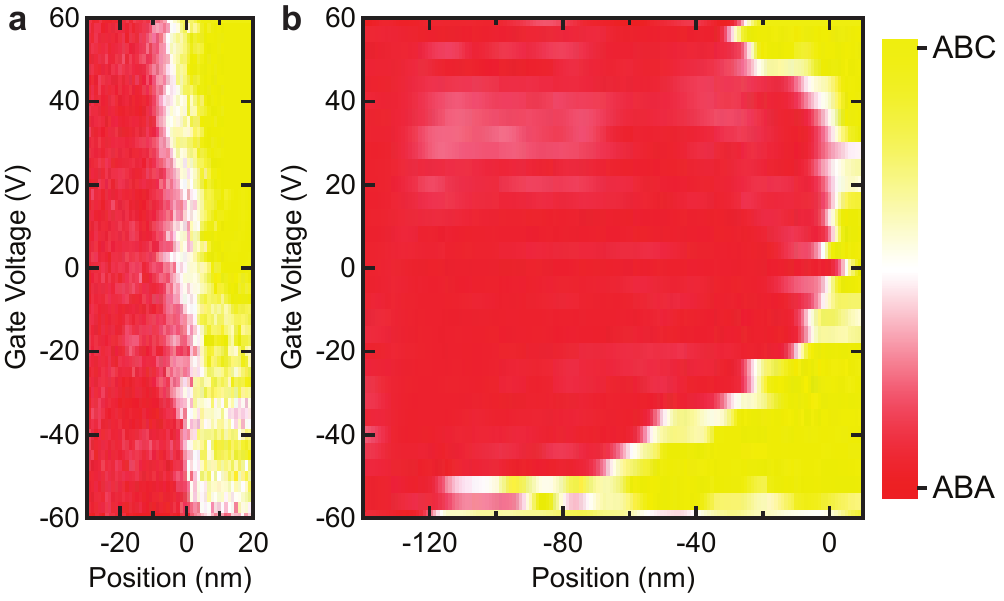} 
\caption{Position of the domain wall as a function of gate voltage. (a) dI/dV spectroscopy as a function of tip position and gate voltage for a pinned soliton.  The data is acquired at a fixed sample voltage of -190 mV.  The position of the soliton does not change with gate voltage.  (b) dI/dV spectroscopy as a function of tip position and gate voltage for a free soliton.  The data is acquired at a fixed sample voltage of -70 mV.  The tip is moving from left (ABA) to right (ABC).  The position of the domain wall is defined by the abrupt jump in the dI/dV trace as a function of the gate voltage.  These specific sample voltages are chosen to best highlight the soliton position for all gate voltages probed; the soliton position is independent of sample voltage.}
\label{fig:gatemovement}
\newpage
\end{figure}

To understand the behavior of the domain wall we take a line cut of the spectroscopy across the boundary at a fixed sample voltage of -150 mV.  Figures ~\ref{fig:linespectroscopy}(c) and (d) show the results for the smooth and abrupt transitions, respectively.  In both cases, the transition from ABA to ABC stacking can be clearly observed at all sample voltages.  Figure ~\ref{fig:gatemovement}(a) shows similar line cuts of dI/dV spectroscopy as a function of gate voltage (and therefore electric field) for the pinned domain wall.  The red (yellow) region corresponds to ABA (ABC) stacking.  We find there is little to no movement of the domain wall in the pinned region as the electric field changes.  However, similar measurements near the center of the trilayer region, where the abrupt transition is observed, show markedly different behavior as a function of gate voltage.  Fig. ~\ref{fig:gatemovement}(b) shows the comparable measurement to Fig. ~\ref{fig:gatemovement}(a) in the unpinned region.  Here, we find that the position of the domain wall remains nearly stable at small gate voltages, but can change by more than 100 nm with the application of large gate voltages.  As the gate voltage (and electric field) becomes larger, more of the sample becomes ABC stacked.  

%%%%%%%%%%%%%%%
%%%%%%%%%%%%%%%%%%

The movement of the ABA/ABC interface can be understood from the energetics of the domain wall.  In the absence of an external STM tip, the domain wall position is determined by pinning, elastic energy and stacking energy (see Supplementary Information for details).  Assuming that stacking shifts occur only parallel to one of the lattice vectors, we obtain a soliton-like profile of the domain wall with a width ranging from $7$ nm for a shear soliton with shifts parallel to the domain wall to $11$ nm for a tensile soliton with shifts perpendicular to the domain wall.  The external STM tip introduces three additional ingredients; an elastic energy for displacing the soliton from its equilibrium position, a repulsive van der Waals potential between the tip and soliton, and an energy imbalance between the ABA and ABC regions under a perpendicular electric field.  The elastic potential for pulling the soliton of length $L$ away from a point $y=\nu L$ (where $\nu$ is the relative position) a distance $d$ is quadratic in $d$, $U_s = \beta_s d^2/[4\nu(1-\nu)L]$ for $d \ll L$, with $\beta_s \simeq 4.5$ eV/nm.  The van der Waals interaction is given by $U_{vdW} = \beta_{vdW}r_0^3/[z^2+2r_0z+(x-d)^2]^{5/2}$, where $\beta_{vdW}=0.05$ eV nm${}^2$ is the repulsion strength, $r_0$ is the radius of curvature of the tip, $z$ is the tip-sample distance, and $x$ is the tip position.  An electric field $E_z$ opens a gap in the ABC region but not in the ABA region which creates a difference in electronic energy \cite{Peeters2010} that depends on the location of the tip.  The induced energy imbalance may be parametrised by a coefficient $\beta_E$ as $U_{E} (x-d) = - \beta_E \int_{x'>d} dx'dy'e^2 E_z^2(x'-x, y'-y)$ where the electric field is computed assuming a spherical tip above the silicon back gate. The integral is taken only over the ABC region ($x'>d$). 

\begin{figure}[h]
\includegraphics[width=5.0cm]{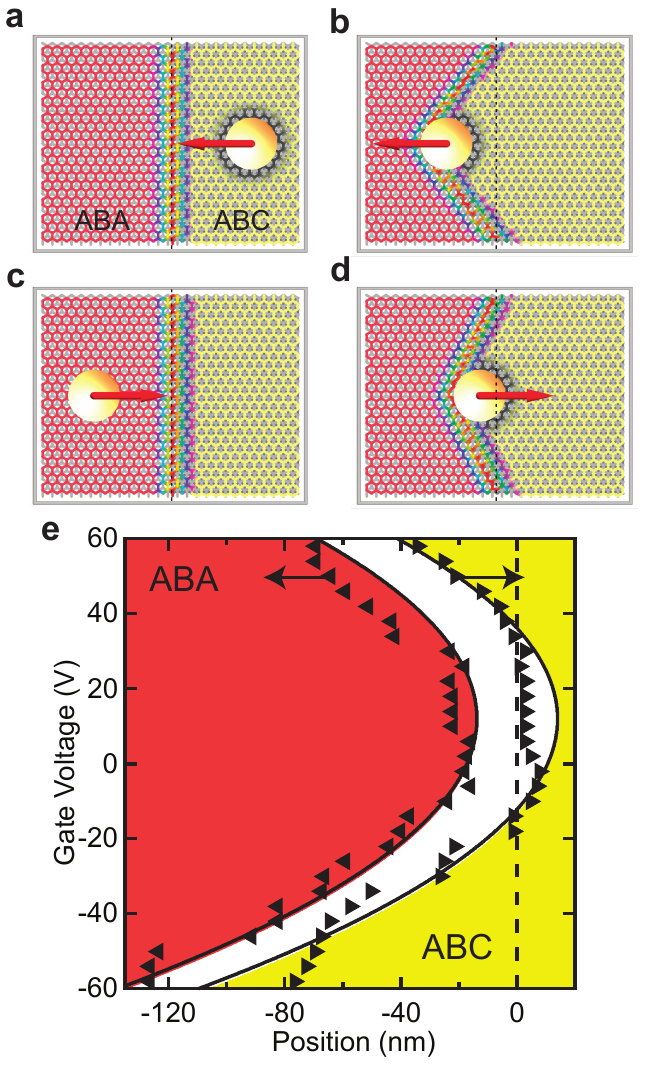} 
\caption{Hysteresis of soliton and modeling.  (a) and (b) Schematics showing the position of the soliton when approaching from the ABC (right) side. The gray circle represents the electronically gapped region under the tip. The soliton is pushed by the approaching tip, both through van der Waals repulsion and through the energetic gain from opening a gap in the ABC region. Pushing proceeds until the tension of the soliton exceeds a threshold, beyond which the soliton jumps back to its relaxed position. (c) and (d) Scanning from the ABA side, the soliton is again repelled by van der Waals, but is attracted by the electronic contribution, since no electronic energy is gained from the ABA side. The snapping threshold, whereupon the soliton jumps towards the left, is thus closer to the ABC region when scanning in this direction.
(e) The snapping position of the soliton as a function of gate voltage for the two different scan directions.  The arrow markers are the experimentally found snapping positions, and they point in the tip scanning direction.  The solid lines are the theoretical fits.  The black dotted line in all panels represents the equilibrium position of the soliton in zero electric field and with no STM tip.  The red and yellow shading represents regions of ABA and ABC stacking in the sample, respectively.  The stacking configuration of the white region depends on the scan direction.}
\label{fig:movement}
\newpage
\end{figure}

Putting all these ingredients together, we obtain the total potential energy of the soliton,
\begin{equation}
U_{\rm tot} = \beta_s d^2/[4\nu(1-\nu)L] + \beta_{vdW}r_0^3/[z^2+2r_0z+(x-d)^2]^{5/2} - \beta_E \int_{x'>d} d^2\mathbf{r}'e^2 E_z^2\, .
\end{equation}
The equilibrium soliton displacement $d_\mathrm{eq}(x)$ is determined by following the local minimum of the potential [$\partial_d U_{\rm tot}(d_\mathrm{eq})=0$] as the tip position $x$ is adiabatically swept. This displacement depends on the tip scan direction and the electric field $E_z$.  It exhibits instabilities beyond certain snapping thresholds, which represent the maximum soliton displacements in a given scan. Scanning from the ABC side towards the ABA side (Figs. ~\ref{fig:movement}(a) and (b)), the tip repels the soliton, which is stretched much like a rubber band. The repulsion is the sum of van der Waals plus the electronic contribution from opening a gap in the ABC region. When the elastic force from stretching the soliton equals this repulsion, the soliton cannot be pushed further.  As the tip continues to move beyond this point, the soliton snaps back towards its original location and the spectroscopy abruptly changes to ABA graphene. Scanning in the opposite direction (Figs. ~\ref{fig:movement}(c) and (d)), the picture is similar, but the electronic contribution is attractive instead of repulsive, and tends to counter the van der Waals repulsion. Hence, the soliton jumps at smaller maximum displacements. In either case, as the soliton jumps under the tip, the measured spectroscopy abruptly changes between ABA- and ABC-type.

%%%%%%%%%%%%%%%
%%%%%%%%%%%%%%%%%%

The rightward-pointing markers in Fig. \ref{fig:movement}(e) map out the experimentally found threshold position as the tip scans from the ABA region into the ABC region, as determined by the location where the topographic signal changes.  These positions are in agreement with the abrupt changes in the dI/dV spectroscopy observed in Fig. ~\ref{fig:gatemovement}(b).  The leftward-pointing markers in Fig. ~\ref{fig:movement}(e) correspond to the opposite scan direction, starting in ABC and moving towards ABA.  The data in both directions can be well fit with our model (solid lines). Parameters $\beta_s \approx 4.5$ eV/nm and $\beta_{vdW}\approx 0.05$ eV nm${}^2$ in our description are predicted by theory, while $L \approx 3~\mathrm{\mu m}$ is determined by Raman spectroscopy and $r_0\approx 250$ nm is based on scanning electron microscope images of similar tips. The tip-sample distance $z=0.5$ nm is set by the tunneling parameters and the traction point $\nu=0.5$ is based on the location of the tip.  The last parameter, $\beta_E$ is constrained by theory to a narrow window (see Supplementary Information). From our fits, we obtain $\beta_E=4.4\times 10^{-4} ~\mathrm{eV}^{-1}$.  As a check for our model, we have repeated the measurement on a different region of the soliton with a second, similar tip.  We find that we can fit this tip's data by only slightly changing the parameters associated with the tip and the traction point (see Supplementary Information).

As with a local electric field created by an STM tip, a global electric field will also move stacking solitons to increase the ABC-stacked area of the device.  This suggests novel and exciting devices that exploit the tunable location of the stacking boundary.  As an example, the soliton may be initially placed underneath one of the source-drain contacts such that the entire conduction path for charge carriers in the device is ABA-stacked.  With the application of a large enough electric field, the soliton will snap into the ABA region, making the device ABC-stacked and gapped, thus quickly turning off conduction in the device.  Such a device would be a good candidate for a graphene FET, offering rapid on-off switching with a high on-off resistivity ratio resulting from the difference in conductivity between ungapped ABA- and gapped ABC-stacked trilayer graphene.

\section*{Methods}
Mechanically exfoliated multilayer graphene was transferred onto high quality single crystals of hBN which were mechanically exfoliated on a SiO$_2$ substrate~\cite{Dean2010}.  Flakes were characterized via  Raman spectroscopy with a WITec Alpha 300RA system using the 532 nm line of a frequency-doubled Nd:YAG laser as the excitation source. The spectra were measured in the backscattering configuration using a 100x objective and either a 600 or 1800 grooves/mm grating.  After depositing the graphene on hBN, Cr/Au electrodes were written using electron beam lithography.  The devices were annealed at 350$^{\circ}$C for 2 hours in a mixture of Argon and Hydrogen and then at 300$^{\circ}$C for 1 hour in air before being transferred to the UHV LT-STM for topographic and spectroscopic measurements.   

All the measurements were performed in UHV at a temperature of 4.5 K. dI/dV measurements were acquired by turning off the feedback circuit and adding a small (5 mV) ac voltage at 563 Hz to the sample voltage. The current was measured by lock-in detection.

\section*{Acknowledgements}
P.S.-J. acknowledges fruitful discussions with J. F. Rossier.

M.Y. and B.J.L. were  supported by the U. S. Army Research Laboratory and the U. S. Army Research Office under contract/grant number W911NF-09-1-0333.  J.I-J.W. was partially supported by a Taiwan Merit Scholarship TMS-094-1-A-001.  J.I-J.W and P.J-H. have been primarily supported by the US DOE, BES Office, Division of Materials Sciences and Engineering under Award DE-SC0001819. Early fabrication feasibility studies were supported by NSF Career Award No. DMR-0845287 and the ONR GATE MURI. This work made use of the MRSEC Shared Experimental Facilities supported by NSF under award No. DMR-0819762 and of Harvard's CNS, supported by NSF under grant No. ECS-0335765.  A.G.B. was supported by the U.S. Army Research Laboratory (ARL) Director's Strategic Initiative program on interfaces in stacked 2D atomic layered materials.  P.S.-J. received financial support from the Spanish Ministry of Economy (MINECO) through Grant no. FIS2011-23713, the European Research Council Advanced Grant (contract 290846) and from the European Commission under the Graphene Flagship (contract CNECT-ICT-604391).

\section*{Supplementary Information}
In this Supplementary Material we present our model for the energetics of a stacking soliton at the interface between ABA- and ABC-stacked 
trilayer graphene. We first consider the elastic energy of a free, relaxed soliton in Sect. \ref{sec:soliton}, 
and how this energy grows under traction in Sect. \ref{sec:pulled}. We then describe  in Sect. \ref{sec:betaE} how 
an electric field $E_z$ affects differently the electronic free energy in ABA- and ABC-stacked trilayer graphene. 
In particular, it opens a gap in the electronic spectrum in the case of ABC- but not ABA-stacking, which results in a lower free energy per unit area in the former with respect to the latter. We show that this energy difference scales as $E_z^2$. In Sect. \ref{sec:Ez} we describe the profile of $E_z$ produced by an idealized tip.  In Sect. \ref{sec:vdW} we model the short ranged van der Waals force between the tip and the soliton, which is present regardless of the backgate potential. 
In Sect. \ref{sec:hysteresis} we describe how to compute the hysteretic evolution of the soliton, in the presence of the elastic, electric and van der Waals forces, as the STM tip scans through its relaxed position in either the ABA-to-ABC or ABC-to-ABA directions.  Finally, in Sect. \ref{sec:fits} we show data and fits for a second tip.

\section{Elastic description of a relaxed soliton \label{sec:soliton}}
In this section we derive, from elasticity theory, the spatial profile, characteristic width and energy density of a relaxed stacking soliton in a graphene bilayer. This description also applies to an ABC/ABA trilayer soliton, assuming that the bottom layer is not strained.

A stacking soliton in a graphene bilayer is a domain wall between an AB- and a BA-stacked region, here taken as $x\to-\infty$ and $x\to\infty$ respectively. A soliton is defined by a interlayer (2D) vector displacement field $\bm{u}(\bm{r})$, with boundary conditions 
\begin{eqnarray}
\bm{u}(x\to-\infty)&=&0\\
\bm{u}(x\to\infty)&=&-\bm{a}_n
\end{eqnarray}
corresponding to AB and BA stacking asymptotics. Here, $n=1,2,3$ denotes the ``flavour'' of the soliton, and $\bm{a}_{1,2,3}$ are the three bond vectors in the uppermost layer, which is the one we will be deforming (we leave the bottom layer fixed for simplicity, without lack of generality). (Note that displacing the top layer of an AB bilayer by a vector $-\bm{a}_n$ transforms it into a BA)

Our aim here is to compute the field $\bm{u}(\bm{r})$ that minimizes the total energy $F=F_u+F_S$, which is the sum of the elastic energy of the deformed top layer, 
\begin{eqnarray}
F_u&=&\frac{1}{2}\int d^2r \left[\lambda \left(\sum_{i}u_{ii}\right)^2+2\mu \sum_{ij}u_{ij}u_{ji}\right] \, , \\
u_{ij}&=&\frac{1}{2}\left(\partial_i u_j+\partial_j u_i\right) \, , 
\end{eqnarray}
plus the stacking energy
\begin{equation}
F_S=\int d^2r V(\bm{u}(\bm{r})) \, .
\end{equation}
This stacking energy $F_S$ is derived from the energy cost of different \emph{uniform} stackings per unit area, $V(\bm{u})$. A uniform AB and BA have minimum stacking energy [$V(\bm{0})=V(-\bm{a}_n)=0$]. Any other stacking has more energy. By incorporating $F_S$ into the total soliton energy, we may arrive at a non-rectilinear soliton profile. Otherwise, the equilibrium soliton has infinite width, to minimize strain in $F_u$. Below, we will also include the \emph{non-uniform} stacking energetics, i.e. the full interlayer shear containing also gradients of $\bm{u}$, to see how the solution is modified.

%\begin{figure}[t] %  figure placement: here, top, bottom, or page
%   \centering
%   \includegraphics[width=0.49\columnwidth]{Vplot.png} 
%   \includegraphics[width=0.49\columnwidth]{Vcut.pdf} 
%   \caption{Taken from Ref. \cite{Alden:PNAS13}: (a) Uniform stacking energy per unit area $V(\bm{u})$. (b) A cut of the potential along an AA-AB-BA-AA sequence.}
%   \label{fig:V}
%\end{figure}

Our model for $V(\bm{u})$ must exhibit the same hexagonal symmetry as the lattice. We will assume $V$ is very large, except along the three crystallographic  $\pm \bm{a}_n$. The cut along these directions takes the form
\[
V(-z\bm{a}_n)=\mathcal{V}(z) \, .
\]
Since any other displacement than the above is energetically prohibitive, this will impose a constraint for the possible soliton displacement fields,
\[
\bm{u}(\bm{r})=-f(\bm{r}) \bm{a}_n \, , 
\]
where $f(\bm{r})$ must be determined, and describes the soliton profile in space. Its boundary conditions are
\begin{eqnarray}
f(x\to-\infty)&=&0 \, , \\
f(x\to\infty)&=&1 \, .
\end{eqnarray}
%We assume, without loss of generality that the chosen interlayer shift is $\bm{a}_n=a\hat{y}$, where $a=0.14$nm is the Carbon-Carbon bond length.

We constrain our soliton ansatz further, by assuming it is a straight ridge, oriented at an angle $\theta$ respect $\bm{a}_n$. Hence
\[
f(\bm{r})=f(\bm{r}\cdot\bm{\hat m}_\theta)
\]
where $\bm{\hat m}_\theta$ is the unit vector normal to the soliton. For concreteness we assume, without loss of generality, that the chosen interlayer shift is $\bm{a}_n=a\hat{y}$, where $a=0.14$ nm is the carbon-carbon bond length. Then, we write \[\bm{\hat m}_\theta=(\cos\theta, \sin\theta).\]

The strain tensor of this soliton reads,
\[
u_{ij}(\bm{r})=-a\left(\begin{array}{cc}
0&\frac{1}{2}\cos\theta\\
\frac{1}{2}\cos\theta&\sin\theta 
\end{array}
\right)f'(\bm{r}\cdot\bm{\hat m}_\theta) \, . 
\]
The associated elastic energy reads
\[
F_u=\frac{a^2}{2}\left(\mu+B\sin^2\theta\right)\int d^2 r\left[f'(\bm{r}\cdot\bm{\hat m}_\theta)\right]^2 \, ,
\]
where $B=\lambda+\mu\approx 12.6~\mathrm{eV/\AA^2}$ is the monolayer bulk modulus, while $\mu\approx 9~\mathrm{eV/\AA^2}$ is half its shear modulus. Note that for a given profile $f(\bm{r}\cdot\bm{\hat m}_\theta)$, the energy of the soliton is mimimum for an orientation $\theta=0$ (i.e. a ``shear" soliton), and maximum for $\theta=\pi/2$ (a ``tensile" soliton).

\begin{figure}[] %  figure placement: here, top, bottom, or page
   \centering
   \includegraphics[width=0.5\columnwidth]{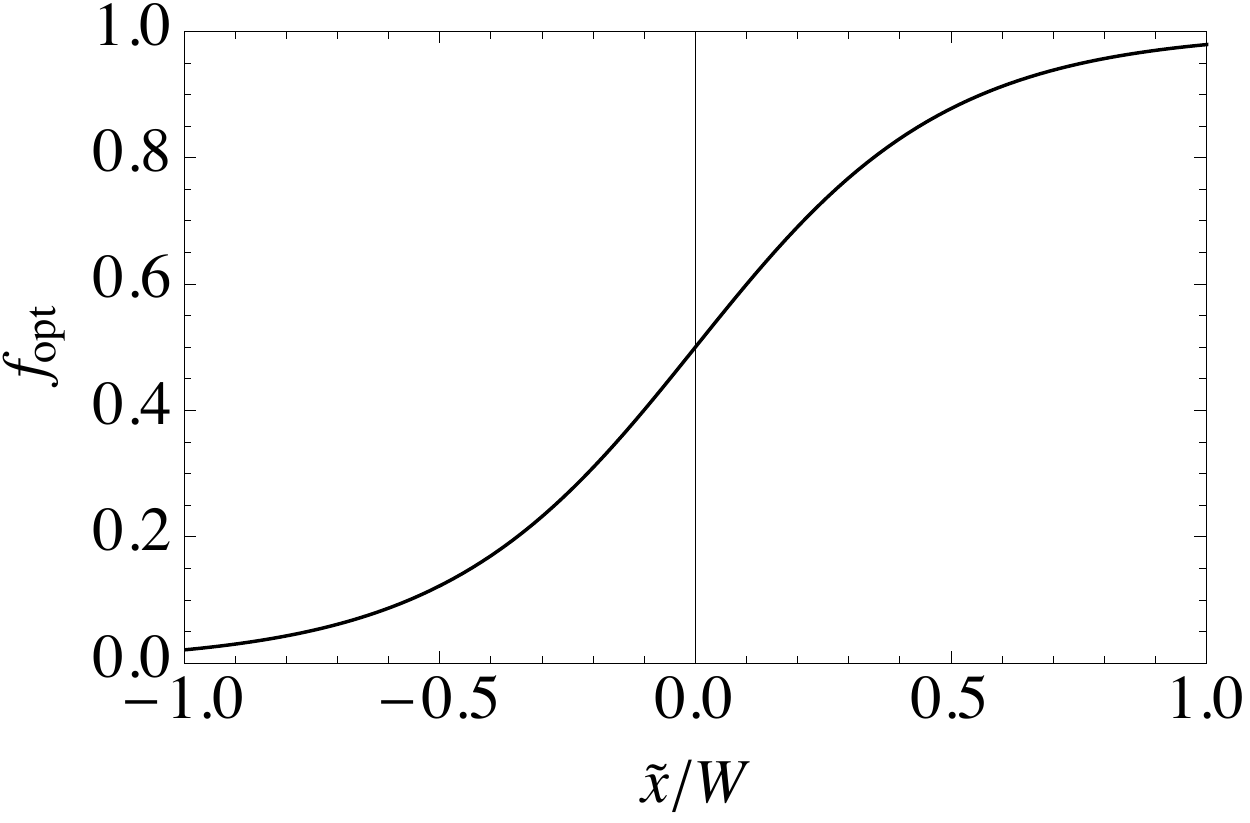} 
   \caption{Soliton profile in terms of length scale $W$ defined in Eq. (\ref{W}).}
   \label{fig:profile}
\end{figure}

We now define coordinates across ($\tilde{x}=\bm{r}\cdot\bm{\hat m}_\theta$) and along ($\tilde{y}=\bm{r}\cdot[\bm{\hat{z}}\times\bm{\hat m}_\theta]$) the soliton.  The profile $f(\tilde{x})$ is obtained by the minimization of the total energy $F=F_u+F_S$. The energy density is independent of $\tilde{y}$, so its integral just gives the length $L$ of the soliton. Hence we are left with
\begin{equation}
F=L\int d\tilde x\left\{\mathcal{V}[f(\tilde x)]+ \frac{a^2}{2}\left(\mu+B\sin^2\theta\right)\left[f'(\tilde x)\right]^2\right\} \, .
\label{Fx}
\end{equation}
A convenient single-parameter model for $\mathcal{V}(z)$ that preserves all relevant symmetries is $\mathcal{V}(z)=\mathcal{V}_0\mathcal{\tilde V}(z)$, with
\[
\mathcal{\tilde V}(z)\approx \left[1-2\cos\left(\frac{\pi}{3} (2z-1)\right)\right]^2
\]
and $\mathcal{V}_0\approx 2\mathrm{meV}/\textrm{atom}=1.57 \mathrm{meV/\AA^2}$ 
\cite{Popov:PRB11}. 
%The energy density $\mathcal{V}_0\approx 3 \mathrm{meV}/\textrm{atom}$, which corresponds to the value at position zero in Fig. \ref{fig:V}(b), will determine the thickness of the soliton profile. 

We can now recast Eq. (\ref{Fx}) in a dimensionless form, 
\begin{equation}
F=\mathcal{V}_0 L W \int_{-\infty}^{\infty} \frac{d\tilde x}{W}\left\{\mathcal{\tilde V}[f(\tilde x)]+ \left[Wf'(\tilde x)\right]^2\right\} \, , 
\label{Fx2}
\end{equation}
where
\begin{equation}
W=\sqrt{\frac{\mu+B\sin^2\theta}{2\mathcal{V}_0}}a \, , 
\label{W}
\end{equation}
is a lengthscale associated to the half-width of the soliton. We obtain $W=7.4$ nm for a shear soliton, and $W=11.6$ nm  for a tensile soliton, in agreement with experiments \cite{Alden2013}.

The solutions that minimize Eq. (\ref{Fx2}) for different $W$ satisfy a scaling invariance $f(\tilde x)=f_\mathrm{opt}(\tilde x/W)$, for some universal function $f_\mathrm{opt}(z)$, so that changing parameter $W$ (for example, adjusting $\mathcal{V}_0$ or $\theta$) just rescales the spread of the relaxed soliton, but not its shape. The function $f_\mathrm{opt}(\tilde x/W)$ can be computed numerically, and is shown in Fig. \ref{fig:profile}. We see that indeed, $W$ is roughly the typical half-width of the soliton. The full width then ranges from 12.2 nm (tensile) to 19.0 nm (shear), in good agreement with experimentally measured values (note that this are roughly twice the full-width-half-maximum values, see Fig. \ref{fig:profile}).

The energy of the soliton solution is 
\begin{equation}
F_\mathrm{opt}=\mathcal{V}_0 L W \int_{-\infty}^{\infty} dz\left\{\mathcal{\tilde V}[f_\mathrm{opt}(z)]+ \left[f_\mathrm{opt}'(z)\right]^2\right\} \, , 
\end{equation}
which equates to an energy per unit length 
\begin{equation}
F_\mathrm{opt}/L\approx 0.649~\mathcal{V}_0 W=0.649\sqrt{\frac{a^2}{2}(\mu+B\sin^2\theta)\mathcal{V}_0} \, , 
%\nonumber\\
%&=&\sqrt{1+1.4\sin^2\theta}\times 105.488 \mathrm{meV/\AA},
\label{Fopt}
\end{equation}
or approximately $F_\mathrm{opt}/L=93.50 \;\mathrm{meV/\AA}$ for a shear soliton, and $F_\mathrm{opt}/L=144.85 \;\mathrm{meV/\AA}$ for a tensile soliton. We see that a tensile soliton has a $55\%$ more energy per unit length than a shear soliton.

\section{Energy of a soliton under traction \label{sec:pulled}}

\begin{figure}[] %  figure placement: here, top, bottom, or page
   \centering
   \includegraphics[width=0.5\columnwidth]{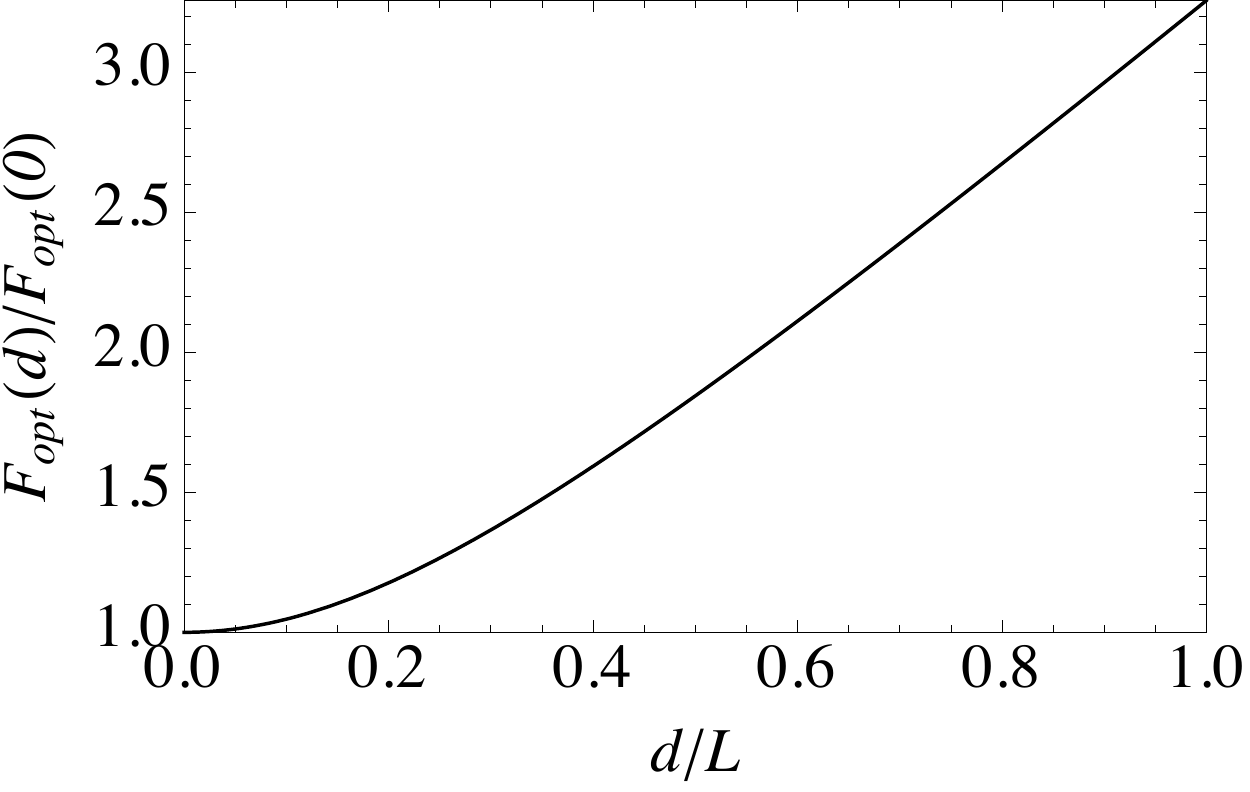} 
   \caption{Energy ratio between a soliton pulled a distance $d$ and the unpulled soliton. $L$ is the length of the soliton.}
   \label{fig:pull}
\end{figure}

We next consider the energetics of a stretched soliton pulled away perpendicularly to its equilibrium direction
by a point like an elastic band. We can generalize the result for a straight soliton Eq. (\ref{Fopt}) to a curved soliton whose radius of curvature is everywhere larger than its width $W$. Then we may approximate
\begin{equation}
F_\mathrm{opt}=F_0\int dL \sqrt{1+\alpha\sin^2\theta}
\label{dFopt}
\end{equation}
with $F_0=0.649\sqrt{a^2\mu\mathcal{V}_0/2}\approx 93.50\,\mathrm{meV/\AA}$ and $\alpha=B/\mu=1.4$. Here $\theta$ is the \emph{local} orientation of the soliton at each point.

In the absence of external traction, the equilibrium shape of the soliton will be a straight line, with an orientation $\theta=0$ everywhere (shear soliton). Assume this is a vertical straight line at $x=0$. If we pull from the center point at $y=0$ a distance $d$ away from $x=0$, the soliton shape will be some function $x_\mathrm{opt}(y)$,  such  that $x_\mathrm{opt}(\pm \infty)=0$ and $x_\mathrm{opt}(0)=d$. The soliton profile 
$x(y)$ minimizes the total energy. From Eq. (\ref{dFopt}) we have
\begin{eqnarray}
F&=&F_0\int dy \sqrt{1+x'(y)^2}\sqrt{1+\alpha\frac{x'(y)^2}{1+x'(y)^2}}\\
&=&F_0\int dy \sqrt{1+(\alpha+1)x'(y)^2} \ .
\end{eqnarray}
The Euler-Lagrange equation for this variational problem is very simple, $x''(y)=0$. Hence, the pulled soliton will remain a straight line to left and right of the pulling point. If the original soliton had a total length $L$, then $x'(y)=2\,\mathrm{sign}(y) d/L$, and the total energy becomes
\begin{equation}
F_\mathrm{opt}(d)=F_0 L\sqrt{1+4(\alpha+1)\left(\frac{d}{L}\right)^2}
\end{equation}
Note that $F_0 L=F_\mathrm{opt}(0)$ is the total energy of the unpulled soliton. The energy ratio between pulled and unpulled solitons is plotted in Fig. \ref{fig:pull}. 
If the pull distance is much smaller than the soliton length $L$, the work done by pulling can be approximated by
\begin{equation}
\Delta F_\mathrm{opt}(d)=F_\mathrm{opt}(d)-F_\mathrm{opt}(0)\approx 2F_0 L(\alpha+1)\left(\frac{d}{L}\right)^2
\label{Fopt2}
\end{equation} 

This equation corresponds to pulling a shear soliton from its center, at $y=L/2$. If the pulling point is generic, at $y=\nu L$, where $0<\nu<1$, and the unpulled soliton is also generic (angle $\theta$) , the above equation generalizes to 
\begin{equation}
\Delta F_\mathrm{opt}(d)\approx 2F_0 L\frac{1}{4\nu(1-\nu)}\frac{1+\alpha}{(1+\alpha\sin^2\theta)^{3/2}}\left(\frac{d}{L}\right)^2\approx \frac{1}{4\nu(1-\nu)} \frac{d^2}{L} \beta_s \, .
\label{Fopt3}
\end{equation}
For a shear soliton, $\beta_s=4.5~\mathrm{eV/nm}$. For a tensile soliton, $\beta_s=1.2~\mathrm{eV/nm}$.

All the above assumes identical energy stacking densities for the AB [$\mathcal{V}(0)$] and BA [$\mathcal{V}(1)$] sides of the soliton, i.e. $\mathcal{V}(0)=\mathcal{V}(1)=0$. While this symmetry is guaranteed by inversion symmetry in a suspended graphene bilayer, it may be broken in a trilayer. In such a system, the ABA stacking energy density has been calculated \cite{Charlier:PRB92,Charlier:C94} to be slightly lower (more stable) than for ABC stacking ($\mathcal{V}(0)>\mathcal{V}(1)$). The relaxed configuration of a soliton pinned at two sites a distance $L$ apart is no longer  a straight line, but becomes bulged towards the ABC side, to minimize the total energy. For realistic parameters, this energy minimum has a curvature, as a function of traction distance around this bulged configuration, that is almost the same as in the case without the ABA/ABC imbalance. We will therefore employ the analytic result Eq. (\ref{Fopt3}) also for a trilayer soliton.

\section{Energetics of graphene trilayer in an electric field}\label{sec:betaE}

%%%%%%%%%%%%%%%%%%%%%%%%%
\begin{figure}
\centering
 \includegraphics[width=0.6\columnwidth]{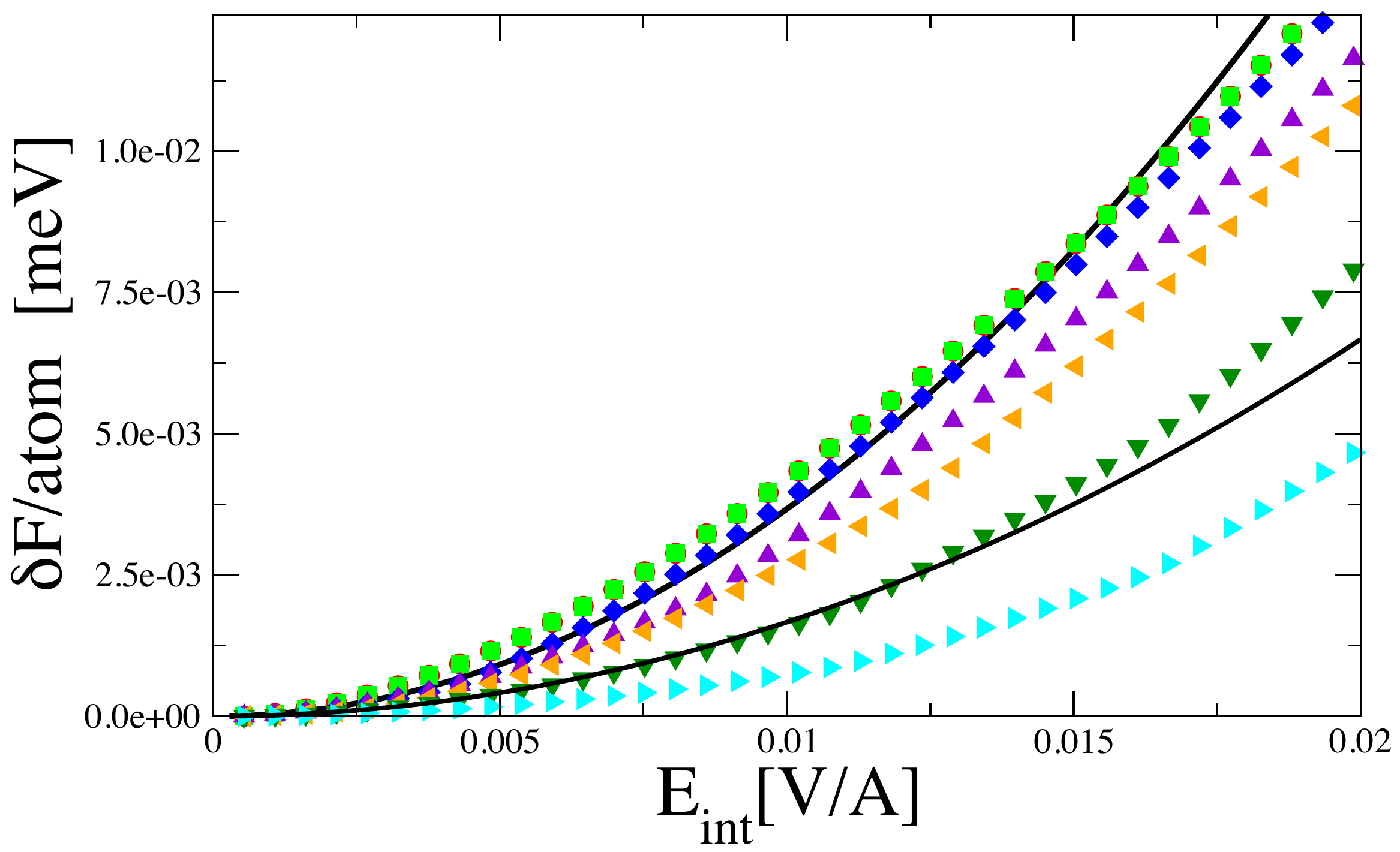} 
\caption{\label{fig:freeenergy} Free energy difference per atom between ABA- and ABC-stacked graphene trilayers,
as a function of the screened, internal electric field. Different set of points correspond to chemical potentials
$\mu = 0$ eV (red), 0.01 eV (light green), 0.02 eV (blue), 0.03 eV (violet) and 0.04 eV (orange),
0.05 eV (dark green) and 0.06 eV (cyan). The solid lines indicate quadratic behaviors with $\beta_E=3 \cdot 10^{-4}$
and $6 \cdot 10^{-4}$ (eV)$^{-1}$.}
\end{figure}

We next consider the energy balance between ABA- and ABC-stacked trilayer graphene in the presence of an electric field. 
Such an electric field arises from the potential energy difference between the sample and an STM tip or a backgate, or both. 
Without field, Ref.~\cite{Charlier:PRB92} explains the dominant ABA stacking in graphite by the presence of a stacking potential favoring
ABA over ABC. ABC is however more sensitive to a perpendicular electric field in that the latter opens a gap in its electronic spectrum, while
at physically relevant field strengths, there is no gap for ABA stacking~\cite{Peeters2010}. For massive two-dimensional 
Dirac fermions with dispersion $\epsilon({\bf k}) = \hbar v_{\rm F} \sqrt{{\bf k}^2 + k_0^2}$, it is a straightforward exercise to show that,
at half filling and low temperature, the difference in free energy  between ungapped ($k_0=0$) and gapped phase is 
$\delta F = F_0-F_{k_0} = 2 \pi \hbar v_{\rm F} k_0^3 \mathcal{A}/3$
favoring the gapped phase, with the sample's area $\mathcal{A}$. It is therefore expected that in the presence of an electric field,
the electronic contribution to the free energy favors ABC stacking over ABA stacking. 

To confirm this expectation, we use the low-energy tight-binding 
Hamiltonians of Refs.~\cite{Peeters2009b,Peeters2010} for ABA and ABC stackings. Energy levels are obtained by exact
diagonalization and the Gibbs free energy is calculated as $F = - k_{\rm B} T \sum_i \ln [ 1+ e^{(\epsilon_i-\mu)/k_{\rm B}T}]$
for both ABA and ABC Hamiltonians, as a function of the electric field $E_{\rm int}$ in the graphene trilayer. 
The latter is incorporated as an on-site energy potential $e E_{\rm int} \, d$ in the top layer and $-e E_{\rm int} \, d$ in the bottom layer, with the
interlayer spacing $d = 3.35$ \AA. Following Ref.~\cite{MacDonald2010}, we take that charge 
screening in the trilayer reduces the externally applied electric field by a factor of 
$\sim8$, $E_{\rm int} \simeq E_{\rm ext}/8$. The geometry of the experiment gives an estimate of 
$E_{\rm ext}\lesssim 0.15$ V/\AA  \, in the experiments, so that the range of interest is roughly $E_{\rm int} \in [0,0.02]$V/\AA. 

Fig.~\ref{fig:freeenergy} shows 
the free energy difference, $\delta F = F_{\rm ABA}-F_{\rm ABC}$, per atom between ABA- and ABC-stacked trilayer graphene for different
chemical potentials, $\mu \in [0,0.06 eV]$. The dependence is quadratic  in the field and systematically favors the ABC phase.
Because the free energy is extensive we write $\delta F = \beta_{\rm E} (e E_{\rm ext})^2 \mathcal{A}$, where we converted
the electric field from internal to external. Taking into account screening~\cite{MacDonald2010} and for
chemical potentials of experimental interest, $\mu \in [-0.05,0.05]$eV, 
we extract from this quadratic behavior $\beta_{\rm E} \in [3 \cdot 10^{-4}, 6 \cdot 10^{-4}]$ (eV)$^{-1}$.
The parameter $\beta_{\rm E}$ depends on $\mu$, which in turn varies slightly with tip and backgate voltages. For the sake of simplicity, and 
because evaluating the experimental value of $\mu$ vs. gate voltage would introduce an additional parameter in the theory, 
we will neglect this latter dependence 
in our
theoretical discussion of the soliton motion, 
and instead consider the bound on $\beta_{\rm E}$ we just extracted.

\section{Electric field profile under a tip\label{sec:Ez}}

A voltage bias applied between the backgate and the tip gives rise to an electric field $E_z(x,y)$ on the sample. In this section we compute this profile, assuming the tip may be modelled by a sphere of radius $r_0$. 

\begin{figure}[] %  figure placement: here, top, bottom, or page
   \centering
   \includegraphics[width=0.8\columnwidth]{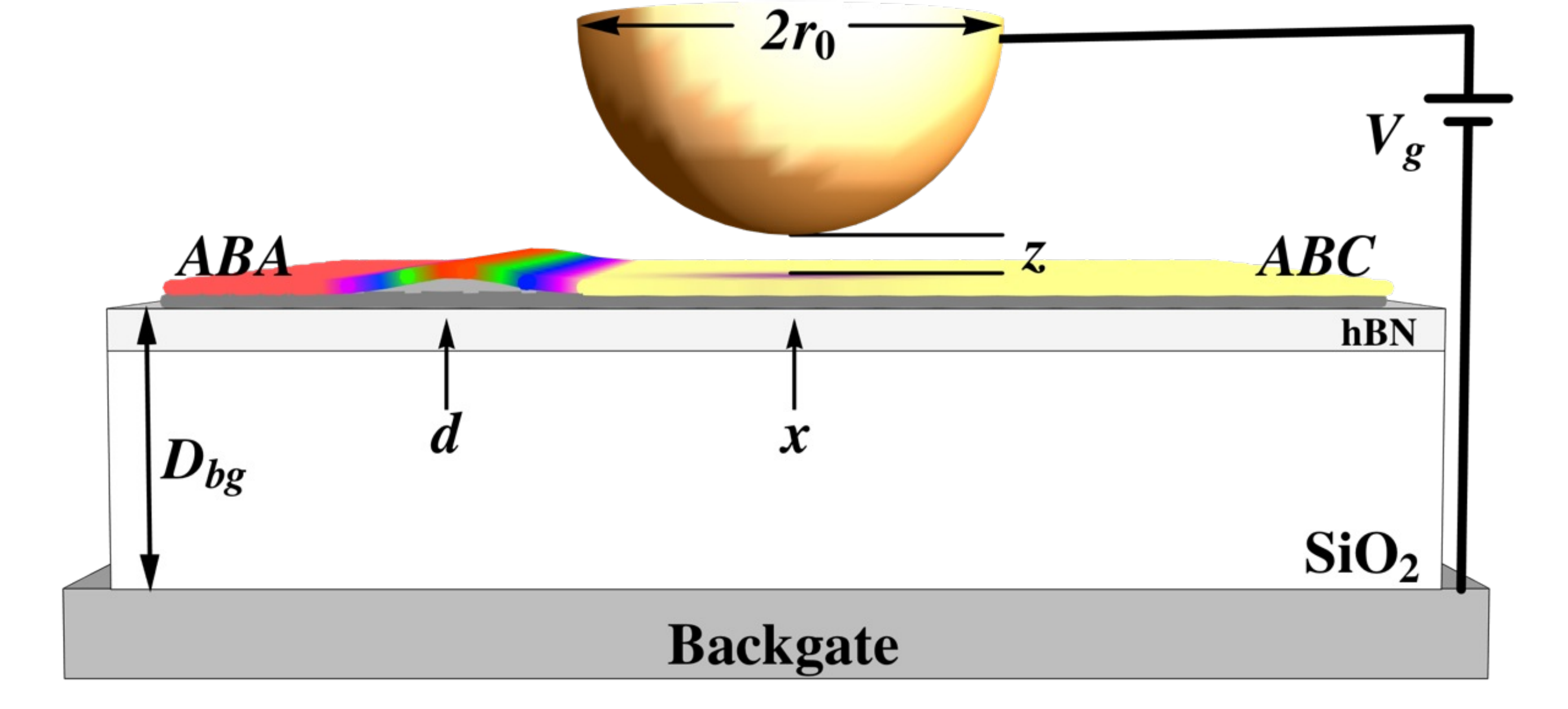} 
   \caption{Sketch of system. A trilayer with a soliton (multicolored boundary between red [ABA] and yellow [ABC]) lies on top of a SiO${}_2$ subtrate, and a hBN layer , of total thickness $D_{bg}=305$ nm. A tip with a radius of curvature $r_0$ hovers at a distance $z$ from the top layer. A bias between tip and backgate creates an electric field that opens a gap in the ABC region.}
   \label{fig:system}
\end{figure}

Consider the setup sketched in Fig. \ref{fig:system}. The sample is sitting on top of a substrate, composed of a $\sim 20$ nm-thick hexagonal Boron Nitride (hBN) crystal immediately under the sample, plus a $\sim 285$ nm-thick layer of SiO${}_2$ below. Both materials have similar dielectric properties, so they will enter the electrostatic problem as a single slab of thickness $D_{bg}=305$ nm with dielectric constant $\epsilon\approx 3.9$. Below it, the backgate is modelled as a flat and infinite metallic plate. The STM tip hovers a
distance $z\approx 0.5$ nm above the sample, which has a thickness $d_T=0.66$ nm. The STM tip's radius of curvature $r_0$ is much larger 
than $z$. Hence, for the purpose of computing the field $E_z$ produced on the sample, it is reasonable to model the tip as a sphere of radius $r_0$. 
Because the sample is very thin we assume that it is transparent and ignore its presence when
computing $E_z$, beyond inducing screening of the electric field as discussed above. 
The problem then reduces to that of a sphere-plate capacitor, see e.g.  Ref. \onlinecite{DallAgnol:RBDEDF09}. The solution takes the form of a set of point charges $Q_n$ and $-Q_n$ at positions $(x,y,Z_n)$ and $(x,y,-Z_n)$, where the origin is chosen on the backgate, and the center of the tip is at $(x,y,Z_0)$. These charges satisfy the recurrence
\begin{eqnarray}
\begin{array}{lcr}
Z_{n+1}=Z_0+r_0^2/(Z_0+Z_{n}) &;\space & Q_{n+1}=Q_n r_0/(Z_0+Z_n)
\end{array}
\end{eqnarray}
The seed position $Z_0=D_{bg}+z+d_T+r_0$ is given by geometry, and seed charge $Q_0=4\pi \epsilon_0 \epsilon V_g R_0$ is fixed by the voltage between the tip and the backgate. Each virtual charge gives a contribution to the field $E_z$ on the sample, which summed up as
\[
E_z(x,y)=eV_g r_0\sum_{n=0}\frac{Q_n}{Q_0}\left(\frac{Z_n-D_{bg}}{\left[x^2+y^2+(Z_n-D_{bg})^2\right]^{3/2}}+\frac{Z_n+D_{bg}}{\left[x^2+y^2+(Z_n+D_{bg})^2\right]^{3/2}}
\right)
\]
This expression allows us to compute the energy gain in ABC, with respect to ABA, in the presence of $E_z$, using $U_E=\beta_E \int dx\,dy\,E_z^2(x,y)$, as discussed in Sect. \ref{sec:betaE}. The number of required images $\pm Q_n$ grows with the ratio $r_0/Z_0$. For the fits in Fig. 4e of the main text, which has $r_0=250$ nm, we have employed 6 images. 
%Likewise, the square of the field, which enters into the ABC-ABA energy imbalance $U_E=\beta_E \int d^2\mathbf{r}' E_z^2(\mathbf{r}')$ discussed in Sect. \ref{sec:betaE}, may be approximated by
%\[
%E_z^2(x,y)\approx(eV_g r_0)^2\sum_{n=0}\frac{Q_n^2}{Q_0^2}\left(\frac{(Z_n-D_{bg})^2}{\left[x^2+y^2+(Z_n-D_{bg})^2\right]^3}+\frac{(Z_n+D_{bg})^2}{\left[x^2+y^2+(Z_n+D_{bg})^2\right]^{3/2}}
%\right)
%\]
%Note that we have neglected all cross terms in the above expression. The reason is that upon integrating $\bm{r}'$ in $U_E$, these lead to special elliptic functions, that are numerically costly. We have checked, however, that dropping them leads only to an approximate rescaling of the $E_z^2(x,y)$ by a factor $0.82$ at $r_0=250$ nm (the value found from fit in Fig. 4e of the main text). This factor was corrected for in said fit. For the main fit in Fig. 4e, the value of $r_0=50$ nm allows us to drop all but the seed charge $Q_0$, since the rest, including cross terms, give a negligible correction in this case.

\section{Van der Waals force between tip and sample \label{sec:vdW}}

London-Van der Waals forces are important players in scanning microscopy, due to the extreme proximity between bulky tips and the sample. The origin of this force is the attraction between instantaneous dipole moments in each of the two bodies. Each pair of dipoles, separated a distance $r$, contributes with an extremely short range potential that is proportional to the mass density $\rho$ in the sample and the tip, $E^{(0)}_{vdW}(r)=-\lambda\rho_\mathrm{tip}\rho_\mathrm{sample}/r^6$, where $\lambda$ is London's constant. Integrating over a spherical tip of radius $r_0$ that hovers at a distance $z$ over the sample (see Fig. \ref{fig:system}), we get an attractive potential with respect to a generic point at a distance $R>r_0$ from the center of the sphere \cite{Hamaker:P37}
\[
E^{(r_0)}_{vdW}(R)=-\frac{4\pi}{3}\frac{\lambda \rho_\mathrm{tip}\rho_\mathrm{sample}r_0^3 }{(R^2-r_0^2)^3} \, . 
\]
If we integrate this over all points in a uniform sample of thickness $d_T\ll R_0$, we obtain the van der Waals attraction between a sphere and a thin plane. The effective van der Waals interaction between the tip and soliton is computed by taking into account that a soliton of width $W$ is expected to have a smaller mass density than the uniform trilayer. The density difference may be estimated as $\Delta \rho=-\frac{1}{3}\rho_\mathrm{sample}a_0/W$, where $a_0=0.24$ nm is the Bravais lattice constant, and the soliton width is $W\approx 7$ nm. It is assumed that only the top layer is strained. 

The difference in van der Waals energy between a trilayer with a soliton, at a distance $x-d$ from the tip, and that of a uniform trilayer, is given by
\[
U_{vdW}(x-d)\approx d_T W\frac{\Delta \rho}{\rho_\mathrm{sample}}\int_{-\infty}^\infty dy E^{(r_0)}_{vdW}\left(\sqrt{(x-d)^2+y^2+(r_0+z)^2}\right)\, , 
\]
where the integral over sample thickness $d_T\approx 0.66$ nm and soliton width $W\approx 7$ nm has been approximated in the limit small $d_T$ and $W$.
The integral over the $y$ coordinate may be evaluated to finally yield
\[
U_{vdW}(x-d)=\beta_{vdW}\frac{r_0^3}{\left[(x-d)^2+2r_0 z+z^2\right]^{5/2}}\, ,
\]
where $\beta_{vdW}=\frac{1}{6}d_T a_0 A$, and $A=\pi^2\lambda\rho_\mathrm{tip}\rho_\mathrm{sample}\approx 1.8$ eV is the Hamaker constant \cite{Hamaker:P37}. This yields $\beta_{vdW}\approx 0.05$ eV nm${}^2$. Note that the resulting van der Waals potential between tip and soliton is repulsive, since the mass density difference $\Delta \rho$ is negative.

\section{Soliton evolution under STM scan \label{sec:hysteresis}}

\begin{figure}
\centering
 \includegraphics[width=0.4\columnwidth]{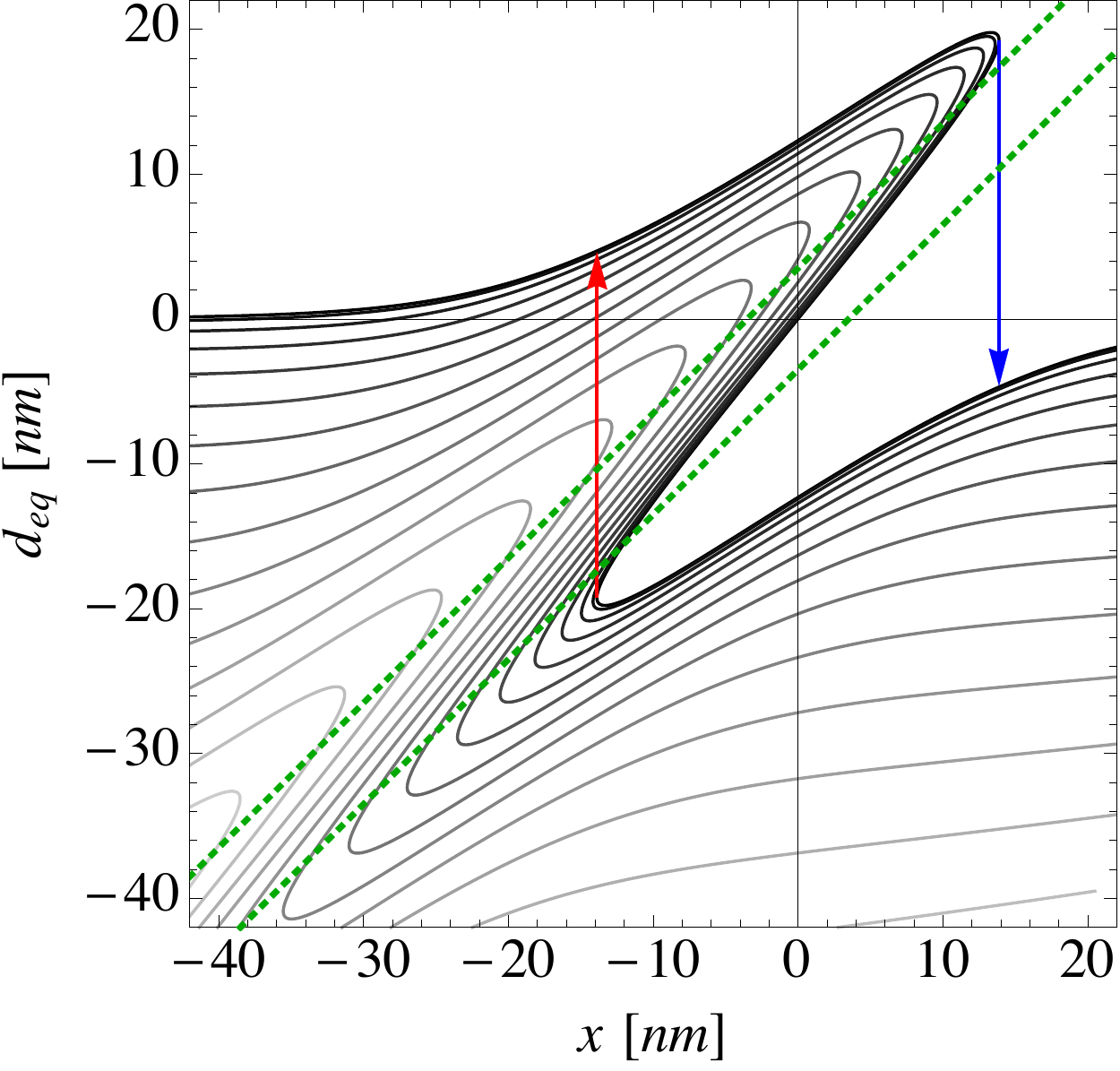} 
\caption{\label{fig:equilibrium} Points of static equilibrium for the soliton displacement $d_\mathrm{eq}$ for different tip positions $x$. Potential parameters as in Fig. 4 of main text. The backgate voltage takes different values, from $V_g=0$ (black curve) to $V_g=70$ V (lightest gray), in steps of 5 V. Soliton snapping thresholds for rightward (ABA to ABC) and leftward (ABC to ABA) tip scans are marked, for $V_g=0$, by blue and red arrows, respectively. The green dotted line denotes the region where the tip lies within the soliton width (i.e. where the differential conductance changes).}
\end{figure}

In this section we discuss the behavior of the soliton as the STM tip is scanned across the sample in the presence of an arbitrary tip-backgate bias. At each tip position $(x,y)$ it exerts a certain force on the soliton, assumed to lie along the $y$ axis in quilibrium. The main source of this tip-soliton interaction comes from the electric field under the tip when the backgate voltage $V_g\neq 0$. The resulting electric field $E_z$ (Sect. \ref{sec:Ez}) creates an electronic energy imbalance between ABC and ABA (Sect. \ref{sec:betaE})
\[
U_E=\beta_E \int_{x'>d} dx'\,dy'\,e^2 E^2_z(x'-x,y'-y),
\]
As a consequence, the tip will repel the soliton when it approaches from the ABC side, but will attract it when coming from the ABA side. The force is proportional to the square of the backgate voltage $V_g$. The value of $\beta_E$, computed in Sect. \ref{sec:betaE}, was found to lie within the range $\beta_E\in[3\cdot 10^{-4},6\cdot 10^{-4}](\mathrm{eV})^{-1}$.

The soliton is also subject to the elastic recovery force (Sect. \ref{sec:pulled}) 
\[
U_s=\beta_s \frac{d^2}{4\nu(1-\nu)L},
\]
where $d$ is the soliton displacement, $L$ is its total length, $y=\nu L$ is the traction point and $\beta_s=4.5$ eV/nm.

Our measurements show, however, that an additional interaction between tip and soliton exists even without a tip-backgate voltage $V_g$, as is clear from the fact that abrupt snapping is observed in the differential conductance for any backgate voltage, including zero. The most natural candidate for this residual force is van der Waals repulsion between the tip and the soliton described in Sect. \ref{sec:vdW}. Such an interaction will push the soliton even for $V_g=0$. Note that other types of interactions \cite{Saint-Jean:JOAP99} could also play a role, but we find that the simple van der Waals model derived in Sect. \ref{sec:vdW} \[
U_{vdW}(x-d)=\beta_{vdW}\frac{r_0^3}{\left[(x-d)^2+2r_0 z+z^2\right]^{5/2}}\,\, ,
\]
is able to correctly reproduce the experimental results. The coupling constant is $\beta_{vdW}\approx 0.05$ eV nm${}^2$.

The total energy is thus $U(d,x)=U_s(d)+U_E(x-d)+U_{vdW}(x-d)$.
To determine the evolution of the soliton displacement $d$ under an adiabatic sweep of the tip position $x$, we compute the equilibrium points, defined by $\partial_d U(d,x)=0$ for different $x$. These are represented in Fig. \ref{fig:equilibrium} for different values of $V_g$.
At $V_g=0$ (no electronic contribution $U_E$, black line), we find that the non-linearity of the model, introduced in particular by the $U_{vdW}$ repulsion, yields a bistable region for a range of positions $x$, corresponding to a displaced soliton behind and in front of the advancing tip. This bistable region ends at snapping thresholds $x\approx \pm 14$ nm (the range of the van der Waals repulsion), where the repelled soliton in front of the tip is too stretched to be pushed further, and snaps behind the tip (arrows in Fig. \ref{fig:equilibrium}). The snapping threshold for opposite scan directions is equal and of opposite sign for $V_g=0$. This represents a hysteretic soliton displacement. At finite $V_g$, the above picture is very similar, but the two hysteretic snapping thresholds are pushed into the ABA region as $-V_g^2$ due to the $U_E$ contribution. The fact that the electronic contribution yields a simple $\propto -V_g^2$ shift is a consequence of the large $r_0\approx 250$ nm, which controls the range of the $U_E$ potential, as compared to the smaller range $\sqrt{2zr_0}\approx 14 $ nm of the van der Waals repulsion $U_{vdW}$. 
The described phenomenology is in quantitative agreement with the experimental results, as shown in Fig. 4e of the main text, and Fig. \ref{fig:movement_si} in the next section.

\section{Movement of soliton \label{sec:fits}}
\begin{figure}
\centering
 \includegraphics[width=0.4\columnwidth]{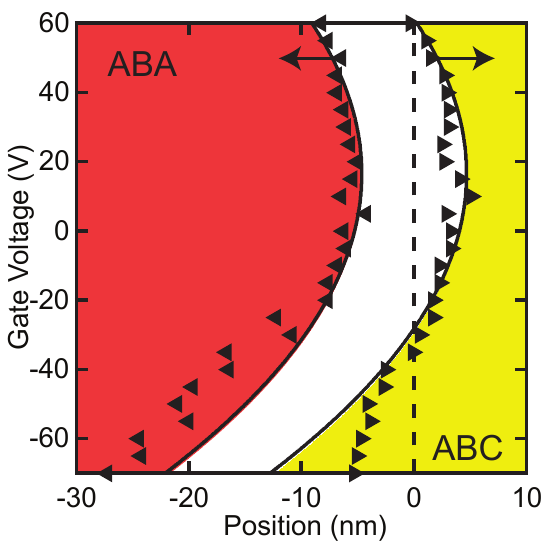} 
\caption{\label{fig:movement_si} Movement of the soliton as a function of gate voltage.  The solid triangles are the experimentally determined locations where the soliton abruptly snaps back to its equilibrium position.  The arrows indicate the scan direction.  The solid black curves are the fits based on our model for the energetics of the soliton.  The red (yellow) areas represent locations which are ABA (ABC) stacked.  The stacking configuration in the white area exhibits hysteresis based on the scanning direction.}
\end{figure}

As a check for our model, we have repeated the measurement of the movement of the soliton as a function of gate voltage for a second tip.  Once again, we have measured the locations where the topography (or spectroscopy) changes for the two different scan directions.  The soliton moves much less in this region but we still observe the abrupt changes in spectroscopy as a function of position. Fig. ~\ref{fig:movement_si} plots the locations where there is an abrupt change for both scan directions.  We are able to fit this data using the same parameters as the main text except for a change in the traction point to within $75~\mathrm{nm}$ of the soliton edge, and slightly reducing $z$ to 0.3 nm.  The change in the traction point is because this set of data was acquired near the bilayer region where the soliton ends and is therefore pinned.  The value obtained in this fit for parameter $\beta_E$, which governs the free energy difference between ABC and ABA under an electric field, is the same, $\beta_E=4.4\times 10^{-4} ~\mathrm{eV}^{-1}$, as for the first tip. This is a relevant consistency check of our model, as $\beta_E$ should be a tip-independent property, intrinsic to trilayer graphene.

\newpage
\end{document}